\documentclass[prd,twocolumn,tightenlines,showpacs,nofootinbib,superscriptaddress,nobalancelastpage]{revtex4-1} 

\usepackage[utf8]{inputenc}
\usepackage{adjustbox}

\usepackage{xfrac}
\usepackage{graphics}     
\usepackage{graphicx}     
\usepackage{longtable}     
\usepackage{url}          
\usepackage{bm}           
\usepackage{natbib} 
\usepackage{placeins}
\usepackage{textpos}
\usepackage{amsfonts,amsmath,amssymb,mathrsfs,bm,amsthm}
\usepackage{float}
\usepackage{booktabs}
\usepackage{array}
\usepackage{tabu}
\usepackage{dcolumn}
\usepackage{rotating}
\usepackage{ulem}
\usepackage{soul}

\newcommand{\RN}[1]{%
  \textup{\uppercase\expandafter{\romannumeral#1}}%
}
\usepackage{longtable}
\usepackage{nicefrac}
\usepackage{amsthm}
\usepackage{color}
\usepackage{cancel}
\usepackage{appendix}
\usepackage{makecell}

\usepackage[colorlinks=true,linkcolor=black,citecolor=blue,urlcolor=blue,bookmarksopen]{hyperref}

\newcommand{\eref}[1]{(\ref{#1})}

\newcommand{\appropto}{\mathrel{\vcenter{
  \offinterlineskip\halign{\hfil$##$\cr
    \propto\cr\noalign{\kern2pt}\sim\cr\noalign{\kern-2pt}}}}}

\begin{document}
\title{Effects of variation of the fine structure constant $\alpha$ and quark mass $m_q$ in  M\"ossbauer nuclear transitions}

\date{\today}

\author{Pavel Fadeev}
\affiliation{Helmholtz Institute Mainz, Johannes Gutenberg University, 55099 Mainz, Germany}

\author{Julian C. Berengut}
\affiliation{School of Physics, University of New South Wales, Sydney, New South Wales 2052, Australia}

\author{Victor V.~Flambaum}
\affiliation{Helmholtz Institute Mainz, Johannes Gutenberg University, 55099 Mainz, Germany}
\affiliation{School of Physics, University of New South Wales, Sydney, New South Wales 2052, Australia}

\begin{abstract}
High accuracy measurements 
in M\"ossbauer transitions open up the possibility to use them in the search for temporal and spatial variation of the fine-structure constant $\alpha$, quark mass $m_q$, and dark matter field which may lead to the variation of $\alpha$ and $m_q$. We calculate the sensitivity of nuclear transitions to variation of $\alpha$ and $m_q$. M\"ossbauer transitions have high sensitivity to variation of quark mass $m_q$ and the strong interaction scale $\Lambda_{QCD}$, to which atomic optical clocks are  not sensitive. The enhancement factors $K$, defined by $\frac{\delta f}{f} =K_{\alpha }\frac{\delta \alpha}{\alpha}$ and $\frac{\delta f}{f} =K_{q }\frac{\delta m_q}{m_q}$ where $f$ is the transition energy, may be large  in some transitions. The 8~eV nuclear clock transition in $^{229}$Th  ($K_q \approx 10^4$) and 76~eV transition in $^{235}$U  ($K_\alpha \approx K_q \approx 10^3$) may be investigated using laser spectroscopy methods.
\end{abstract}

\maketitle


\section{Introduction}
M\"ossbauer spectroscopy has been used for diverse purposes ranging from gravitational redshift of light \cite{Pound1960} to determinations of solids, atomic, and nuclear properties. The sensitivity of M\"ossbauer transitions can reach the $10^{-18}$ level, see e.g.  \cite{GRS}. Moreover, a recent paper \cite{Gratta2020} claims that for a variable perturbation the sensitivity may reach $\Delta E \sim 10^{-15} -10^{-17}$ eV which corresponds to a $10^{-20}-10^{-22}$ relative sensitivity to the frequency shift. For comparison, the best atomic clock limit on relative changes of $\alpha$ is $1.0(1.1) \times 10^{-18}$ per year \cite{Lange2021} (atomic transition frequencies depend on $\alpha$ due to the relativistic effects \cite{Dzuba1999PRL,Dzuba1999PRA,FlambaumDzuba2009}) while the limit on variation of $m_q$ is $0.71(44) \times 10^{-14}$ per year
\cite{Guena} (here sensitivity to $m_q$ comes from the nuclear magnetic moments in Cs and Rb hyperfine transitions \cite{Tedesco}). High  sensitivity motivates the study of nuclear transitions for topics of fundamental physics, such as variation of the fundamental constants \cite{Flambaum2006,Fadeev2020}, search for new particles and interactions \cite{Fadeev2020,Gratta2020}, and search for dark matter (see below). Different possibilities to produce a nucleus in upper state of    M\"ossbauer transition are discussed e.g. in Ref. \cite{Gratta2020}.

\subsection{Variation of the fine structure constant $\alpha$}

The search for temporal and spatial variation of the fine structure constant $\alpha$ is an ongoing interdisciplinary endeavour spanning the fields of astrophysics, molecular, atomic, nuclear, and solid state physics \cite{review2017, Safronova2018}. 
We elucidate the usage of high precision M\"ossbauer spectroscopy in the search for variation of  $\alpha$. The sensitivity to the change of $\alpha$ is encoded in the enhancement factor $K_{\alpha }$, defined as
\begin{align}
  \frac{\delta f}{f} =K_{\alpha } \frac{\delta \alpha}{\alpha} \, .
\end{align}
A nuclear transition energy $f$ would change by $\delta f$ due to a change of $\alpha$ by  $\delta \alpha$. Values of $K_{\alpha }$ for current atomic clocks are of the order 0.1--10  \cite{Dzuba1999PRL,Dzuba1999PRA,Huntemann2014,FlambaumDzuba2009,DFS2018,Safronova2019}. In nuclei $K_{\alpha }$ may be found from the following relation \cite{Flambaum2006}:
\begin{align}\label{Kalpha}
K_{\alpha } = \Delta E_{\text{C}}/f \, ,
\end{align}
where $\Delta E_C$ is the change in Coulomb energy in this transition.
 In the $^{\textrm{229}}\textrm{Th}$ 8 eV nuclear clock transition studied e.g. in Refs.  \cite{PeikTamm2003,Reich1990,Helmer1994,Beck2007,Beck2009,Tkalya2015,Minkov2017,
 Porsev2010,Campbell2012,Wense2018,Thirolf2019,Tkalya2020,Nickerson2020,Porsev2010a,Porsev2010b,Dzyublik2020,Seiferle2019,Masuda2019,Yamaguchi2019,Sikorsky2020} and expected to be a highly sensitive probe for time variation in $\alpha$ \cite{Flambaum2006,Wiringa,Auerbach,Litvinova09,Thirolf2019b,Julian2010,Beeks21}, our recent analysis \cite{Fadeev2020} gives $K_{\alpha }$ of $10^4$.

\subsection{Variation of the quark mass and strong interaction} 
 
Another dimensionless parameter which affects nuclear transition energies is $X_q \equiv m_q/\Lambda_{QCD}$, where $m_q=(m_u+m_d)/2$ is the quark mass and $\Lambda_{QCD}$ is the QCD  scale. 
Here we measure $m_q$ in units of $\Lambda_{QCD}$, i.e. in the calculations we keep  $\Lambda_{QCD}$ constant.
The energy of a nuclear transition  may be presented as 
\begin{align} \label{fs}
    f= \Delta E_{\text{C}} +  E_S \, ,
\end{align}
where $ E_S$ is the difference in bulk binding energies of the excited and ground states (including kinetic and strong potential
energy but excluding the Coulomb interaction energy). Thus, using experimental value of the transition energy $f$ and calculated value of the Coulomb energy difference $\Delta E_{\text{C}}$, we can find $E_S=f - \Delta E_{\text{C}}$.
The dependence of $ E_S$ on quark mass was calculated in Ref. \cite{Wiringa}:
\begin{align} \label{deS}
\frac{\delta E_S}{E_S}= -1.45 \frac{\delta m_q}{m_q} \,.
\end{align}
Using Eqs. (\ref{Kalpha} -- \ref{deS}), we obtain 
\begin{align}\label{Kq}
  \frac{\delta f}{f} =K_{q} \frac{\delta m_q}{m_q}, \,\,\,\,  K_q=1.45 (K_{\alpha }-1).
\end{align}

\section{Possible physical origins of $\alpha$ and $m_q$ variation in  M\"ossbauer transitions}

There are several possible physical origins of $\alpha$ and $m_q$ variation in  M\"ossbauer transitions, some of which we illuminate here. Many popular theories extending the Standard Model contain scalar fields $\phi$ which interact with quarks $q$ as $-(\phi/\Lambda_q)  m_q \bar q q$. Here $\Lambda_q$ is the interaction constant. This interaction may be added to the mass term in the Lagrangian $-m_q \phi \bar q q$ and presented as a dependence of the effective quark mass $m_q(\phi)=m_q (1 + (\phi/\Lambda_q))$ on the field $\phi$ (see e.g. \cite{Stadnik,Dy}). Another possibility is that an interaction $(\phi/4 \Lambda_{\gamma}) F_{\mu \nu} F^{\mu \nu}$ between the scalar field and electromagnetic field $F^{\mu \nu}$ may be added to the electromagnetic term in the Lagrangian $F^{\mu \nu}F_{\mu \nu}/4$. This will manifest as a dependence of the fine structure constant $\alpha(\phi)=\alpha (1 + (\phi/\Lambda_\gamma))$ on the field $\phi$ (see e.g. \cite{Arvanitaki,Stadnik,Dy}). 
Assuming that the source and absorber of the  M\"ossbauer radiation are separated by some distance  $r$, the values of $\alpha$ and $m_q$ can be different at these points, if the field $\phi$ varies in space (see below).

\subsection{Yukawa field $\phi$}

The field  $\phi$ may vary since the  interaction between the field $\phi$ and Standard Model particles leads to the Yukawa field $\phi = C \exp(-mr)/r$ produced by any massive body. The coefficient C has been calculated in Ref.~\cite{Dy}. In this way the presence of a massive body affects the fundamental constants. 

For example, in the experiment \cite{Dy} variation of the field $\phi$ and $\alpha(\phi)$ was produced by moving a 300~kg lead mass back and forth, affecting the ratio of the transition frequencies in Dy and Cs atoms. These have different dependence on $\alpha$, since in Dy $K_{\alpha}$ is strongly enhanced \cite{Dzuba1999PRL,Dzuba1999PRA,FlambaumDzuba2009}. In the case of  M\"ossbauer transitions, a mass may perform oscillating motion toward emitter (or absorber) of the radiation, producing a difference in the transition frequencies between the emitter and absorber
$\delta f = f (K_{\alpha} \delta \alpha/\alpha + K_{q} \delta m_q/m_q)$ 
which oscillates  with the frequency of the mass motion. 
  
Alternatively, the Yukawa field $\phi$  may be generated on a microscopic scale. In a recent paper \cite{Gratta2020} a technique to search for new scalar and tensor interactions at the submicrometer scale is presented.
They suggest to place the optically flat ``attractor" (source of Yukawa field $\phi$), which perturbs the M\"ossbauer absorber frequency, on a micropositioner. This arrangement will provide a high sensitivity to the field $\phi$ with mass corresponding to the submicron Compton wavelength. Importantly, the paper \cite{Gratta2020} provides estimates of the systematic effects produced by the electromagnetic interactions and concludes that they are very small: the estimated sensitivity is  $\Delta E \sim 10^{-15}$ -- $10^{-17}$ eV which corresponds to  $\delta f/ f \sim 10^{-20}$ -- $10^{-22}$. 
Based on these estimates and using  $\delta f = f (K_{\alpha} \delta \alpha/\alpha + K_{q} \delta m_q/m_q)$ with values of $K$ from Table \ref{tab:moreelements}, we obtain sensitivity to the  variations $\delta \alpha/\alpha \sim \delta m_q /m_q \sim 10^{-20}$ -- $10^{-23}$. This estimate may be optimistic but we should compare it with the current limits from atomic transitions $\delta \alpha/\alpha \sim 10^{-17}$ -- $10^{-18}$ and  $\delta m_q/m_q \sim 10^{-14}$.

A gradient of $\phi$ may also be due to the Yukawa field produced by a nearby mountain or by the whole Earth if this field has a large range (a small mass). Here the situation is somewhat similar to the measurements of the gradients of the gravitational field. 

\subsection{Dark matter field $\phi$}

If we identify the scalar field with dark matter, a gradient of the field $\phi=\phi_0 \cos(\mathbf{k}\cdot\mathbf{r} -\omega t)$ appears due to the non-zero wave vector $k \approx m v /\hbar$, where $v$ is the speed of Earth in the reference frame of the Galaxy and $\omega \approx mc^2/\hbar$, $\phi_0$ is determined by the dark matter mass density (see e.g. \cite{Stadnik,Arvanitaki}). In this case we have oscillating  $m_q(\phi)=m_q (1 + (\phi/\Lambda_q))$ and $\alpha(\phi)=\alpha (1 + (\phi/\Lambda_\gamma))$, which depend on the position ${\bf r}$. Therefore, the  dark matter field $\phi$ induces oscillations in the difference of the transition energies between separated emitter and absorber of the M\"ossbauer radiation.

A gradient of the field $\phi$ may exist in the transient field of passing clumps of dark matter, Bose stars, domain walls, etc. A gradient of $\phi$ may also exist in the field of scalar particles captured by Earth  (see e.g. reviews \cite{Safronova2018,halo} and references therein).

\subsection{Comparison of transition frequencies which have different dependence on fundamental constants}

Search for variation of the fundamental constants in atomic experiments has been done using time dependence measurement of the ratio of two transition frequencies  which have different dependence on the fundamental constants.
A M\"ossbauer transition might be compared with a transition of approximately the same frequency in a highly charged ion. It may be challenging to find such ion transition, but they may be sought in the spectra of ions with open f-shell, which are very dense. 

In the case  of the 8~eV nuclear clock transition in $^{229}$Th, laser optical spectroscopy methods, such as frequency comb, may be used  for  comparison with other transitions. High frequency sources of coherent radiation, based on the multiplication of the frequencies of the laser field,  should allow one to extend this approach to 76 eV transition in $^{235}$U. 

\section{Calculation of the sensitivity to $\alpha$ and $m_q$  variation in nuclear transitions}

To deduce $K_\alpha$ for a particular transition, $\Delta E_C$ must be calculated. This can be done using measurements of the changes in the mean square  charge radius $\Delta \langle r^{2}\rangle$ and intrinsic quadrupole moment $\Delta Q_0$ between the ground and excited states \cite{Julian2009,Fadeev2020}:
\begin{align} \label{derivativeEc}
\Delta E_{\text{C}}=
\langle r^{2}\rangle \frac{\partial E_{\text{C}}}{\partial \langle r^{2}\rangle}
\frac{\Delta\langle r^{2}\rangle}{\langle r^{2}\rangle}+
Q_{0}\frac{\partial E_{\text{C}}}{\partial Q_{0}}
\frac{\Delta Q_0}{Q_{0}} \, .
\end{align}
To extract values of intrinsic electric quadrupole moments $Q_0$ from experimental  data for the electric quadrupole moments $Q_{lab}$ we use the following relation for rotating deformed nuclei: 
\begin{align}
\label{Qlab}
    Q_{lab} =  Z Q_0 \frac{I (2I-1)}{(I+1)(2I+3)} \, .
\end{align}
The use of this formula in nuclei with a small or zero deformation is not justified, however the electric quadrupole in such nuclei is small and has little effect on the final result.

To calculate derivatives $\frac{\partial E_{\text{C}}}{\partial \langle r^{2}\rangle}$ and  $\frac{\partial E_{\text{C}}}{\partial Q_{0}}$ we model the nucleus as a spheroid \cite{Julian2009,Fadeev2020}. In such a model
\begin{align} \label{Ec}
E_{\mathrm{C}} &= E_{\mathrm{C}}^{0} \, B_{\mathrm{C}} \, , \\
E_{\mathrm{C}}^{0} &=\frac{3}{5} \frac{q_e^{2} Z^{2}}{R_{0}} \, ,
\end{align}
where $q_e$ is the electron charge, $Z$ is the number of protons, and for a prolate spheroid $\left( Q > 0 \right)$:
\begin{align} 
  \label{Bcoul}
B_{\text{C}} &= \frac{(1-e^2)^{1/3}}{2e} \textrm{ln} \left( \frac{1+e}{1-e} \right)  \, ,
\end{align}
with $e$ being the eccentricity. For an oblate spheroid $\left( Q < 0 \right)$, with the eccentricity defined such that it stays positive:
\begin{align} 
  \label{Bcoul}
B_{\text{C}} &= \frac{(1+e^2)^{1/3}}{e} \arctan \left( e \right)  \, .
\end{align}

If one of the $\Delta Q_0$ or $\Delta \langle r^2 \rangle$ measurements is missing, one way to estimate the result is by using the ansatz of constant charge density between isomers, which is equivalent to the ansatz of constant volume \cite{Fadeev2020}. In such a case, for a spheroid,
\begin{equation}
\label{eq:dQdr2}
\frac{dQ_0}{d\langle r^2 \rangle} = 1 + \frac{2\langle r^2 \rangle}{Q_0} \, .
\end{equation}
Note that in Refs. \cite{Julian2009,Fadeev2020} we tested the accuracy of Eq. (\ref{derivativeEc}) and constant density anzats Eq. (\ref{eq:dQdr2}) using results of  Hartree-Fock-Bogolyubov calculations \cite{Litvinova09} of  $\Delta E_C$, $\Delta Q_0$ and $\Delta \langle r^{2}\rangle$ for the $^{229}$Th  nuclear transition. We estimated the error in the constant density ansatz of $\sim 25\%$.

In Table \ref{tab:moreelements} we compile an extensive list of $\Delta E_C$ and enhancement factors $K$ for M\"ossbauer transitions. The measured values of $\Delta \langle r^{2}\rangle$  and $Q$, which we use as an input, are presented in Appendix B. The accuracy of the electric quadrupole moments measurements at the moment  is insufficient for extraction of reliable values $\Delta Q_0$. Therefore, we base our results on the measured values of $\Delta \langle r^{2}\rangle$ (which in any case gives the main contribution to $\Delta E_C$)  and constant density ansatz Eq.~\eref{eq:dQdr2} to find  $\Delta Q_0$. Note that if we neglect $\Delta Q_0$, the value of $\Delta E_C$ would increase.
Therefore, the constant density ansatz gives us a conservative estimate of $\Delta E_C$ and $K$.

For some elements $\Delta \langle r^{2}\rangle$ is absent in the literature, to the best of our knowledge. For an estimate, we could  use the constant density ansatz  Eq. (\ref{eq:dQdr2}) to find $\Delta \langle r^{2}\rangle$  using  known values of the electric quadrupole moments $Q$ in 
\mbox{$^{235}$U 46\,keV},
\mbox{$^{233}$U 40\,keV},
\mbox{$^{179}$Hf 123\,keV}, 
\mbox{$^{165}$Ho 95\,keV}, 
\mbox{$^{160}$Ho 60\,keV}, 
and \mbox{$^{158}$Ho 67.2\,keV}.
However, the errors  in $\Delta Q_0$ and $K_{\alpha}$ are too large in these cases, so we can not make definite predictions.

We see in Table \ref{tab:moreelements} that the average value of $|\Delta E_C|$ in medium and heavy deformed nuclei is $\sim$ 70 keV. Therefore, we may assume $\Delta E_C \sim$ 70 keV and $|K_{\alpha}| \sim 70 $ keV$/f$ in all medium and heavy deformed nuclei where accurate data for  $\Delta \langle r^{2}\rangle$ are not available. In light nuclei and spherical nuclei $|\Delta E_C| \sim 30$ keV .

Two exceptional transitions presented in Table~\ref{tab:moreelements} are the 8~eV nuclear clock transition in $^{229}$Th and the 76~eV transition in $^{235}$U.
Investigation of the $^{229}$Th transition using laser spectroscopy methods has long been discussed in the literature, however new sources of coherent radiation cover the range up to 100 eV (see e.g. \cite{UVlaser}), so 76 eV transition in $^{235}$U may be investigated using high precision spectroscopy too. The probability of the photon emission in the bare $^{235}$U nucleus is very small but it is significantly enhanced by the electronic bridge mechanism in many-electron ions \cite{Berengut2018} (see also \cite{Hinneburg}). To avoid discharge of the 76 eV nuclear excited state by electron emission, ionization potential of the uranium ion should exceed 76 eV. This condition is satisfied in ions with charge bigger than 6. The values of the enhancement factors for 76 eV transition in $^{235}$U,  $K_q\approx K_{\alpha} \approx 10^3$, are estimated in the Appendix A using the Nilsson model.

In summary, we show that nuclear transitions are a sensitive tool in the search for the variation of the fine structure constant $\alpha$  and especially  variation of the strong interaction parameter $m_q/\Lambda_{QCD}$ to which atomic optical transitions are not sensitive. We calculate the sensitivity to these parameters, presented as the enhancement factors $K_{\alpha}$ and $K_q$, for a number of M\"ossbauer transitions, 8 eV transition in $^{229}$Th, and 76 eV transition in  $^{235}$U . 

\textit{Acknowledgements} --- We are grateful to S. Bladwell who participated at the initial stage of this work. This work was supported by the Australian Research Council Grants No. DP190100974 and DP200100150 and the Gutenberg Fellowship.

\begin{table*}[tbp]
\caption{Sensitivity of M\"ossbauer transitions to variation of the fine structure constant and of the quark  mass. Coulomb energy shifts $\Delta E_{\text{C}}$ and enhancement factors $K$ calculated using data which we list in the Appendix.
We present in the Table the ``experimental'' errors which are determined from the errors in $\Delta \langle r^2 \rangle$  values. 
Our estimate for the constant density ansatz ``theoretical'' error is $\sim$ 25 \%. 
}
\centering
\begin{ruledtabular}
\begin{tabular}{lccccccccc}
 & \multicolumn{1}{c}{$T_{1/2}$} &  \multicolumn{2}{c}{$j \pi$} & \multicolumn{1}{c}{$\Delta E_{\text{C}}$ (keV)} & \multicolumn{1}{c}{$K_{\alpha}$} &  \multicolumn{1}{c}{$K_{q}$} \\
 & excited state& gr & ex & const. density  & const. density  &  const. density   \\
\hline \\
  $^{57}$Fe
\, \, 14.4 keV & 98 ns &
$1/2-$  & $3/2-$  
& 39($9\%$) & $2.7 (9\%)$  &
$2.4(14 \%)$
\\ \\
$^{67}$Zn
\, \, 93.3 keV & 9.07 $\mu$s &
$5/2-$  & $1/2-$ 
& $-35 (27\%)$ & $-0.37 (27\%)$ & $-1.99 (7\%)$
 \\ \\
$^{83}$Kr
\, \, 9.3 keV & 147 ns &
$9/2+$  & $7/2+$ 
  & $-15\,(25\%)$ &
  $-1.6\,(25\%)$ &
$-3.8(15\%)$ 
  \\ \\
   $^{99}$Ru
\, \, 90 keV & 20.5 ns &
$5/2+$  & $3/2+$  
  & $-59\,(26\%)$ &
  $-0.66\,(26\%)$ &
$-2.40 (10\%)$
  \\ \\
  $^{119}$Sn
\, \, 23.9 keV & 17.8 ns &
$1/2+$  & $3/2+$   
&  $-25.1(3\%)$ & $-1.053(3\%)$ & $-2.98 (1\%)$
 \\ \\
   $^{121}$Sb
\, \, 37.2 keV & 3.5 ns &
5/2+ & 7/2+ 
& $183 $ & $4.91$ & $5.67$  \\ \\
  $^{125}$Te
\, \, 35.5 keV & 1.48 ns &
1/2+ & 3/2+ 
& $-13.3 (17\%)$ & $-0.37 (17\%)$ & $-1.99(5\%)$ \\ \\
$^{127}$I
\, \, 57.6 keV & 1.95 ns 
& $5/2+$  & $7/2+$  &
56.1 & 0.97 & $-0.04$\\ \\
  $^{129}$I
\, \, 27.8 keV & 16.8 ns &
7/2+ & 5/2+ 
& $ -69.7$ & $-2.51$ & $-5.08$ \\ \\
 $^{149}$Sm
\, \, 22.5 keV & 7.6 ns &
$7/2-$  & $5/2-$ 
  & $-5.3\,(29\%)$ &
  $-0.24\,(29\%)$ &
$-1.79 (6\%)$
   \\ \\
$^{151}$Eu \, \, 22 keV & 9.5 ns &
5/2+ &7/2+
& $-99\,(29\%)$ & $-4.6\,(29\%)$ &  $-8.1 (24\%)$   \\ \\
$^{153}$Eu
\, \,  83.4 keV & 0.80 ns &
$5/2+$  & $7/2+$  
  & $10.2\,(25\%)$ &
  $0.12\,(25\%)$ &
 $-1.27 (3\%)$ \\ \\
$^{153}$Eu \, \, 103 keV & 3.9 ns &
5/2+ & 3/2+
& $321\,(15\%$)  & 3.1\,(15\%) & 3.1(23\%)   \\ \\
$^{155}$Gd
  \, \, 86.5 keV & 6.35 ns &
  $3/2-$ & $5/2+$ 
  & $ 22\,(25\%)$ &
  $0.25\,(25\%)$ &
$-1.09 (8\%)$
  \\ \\
$^{155}$Gd \, \, 105 keV & 1.18 ns &
$3/2-$ & 3/2+ 
& 30 (25\%)  & 0.28\,(25\%)  & $-1.04 (10\%)$   \\ \\
$^{157}$Gd
 \, \, 64 keV & 0.46 ms &
 3/2$-$ & 5/2+ 
& $-55\,(25\%)$ & $-0.86\,(25\%)$ & $-2.69(12\%)$  \\ \\
$^{161}$Dy
\, \, 25.7 keV & 29 ns & 
5/2+ & $5/2-$ 
  & $-29\,(25\%)$ &
  $-1.14\,(25\%)$ &
$-3.10 (13\%)$ 
   \\ \\  
$^{161}$Dy
\, \, 43.8 keV & 0.78 ns &
5/2+ & 7/2+ 
  & $6.3\,(25\%)$ &
  $0.14\,(25\%)$ &
 $-1.24(4\%)$ \\ \\
$^{161}$Dy
 \, \, 75 keV  &  3.2 ns &
 5/2+ & 3/2$-$ 
 & $-31\,(25\%)$ &  $-0.42\,(25\%)$ & $-2.06 (7\%)$   \\ \\
  $^{181}$Ta
\, \,  6 keV & 6.05 ms &
7/2+ & 9/2$-$ 
 & 191\,($25\%$) & 30\,($25\%$) & 43 ($26\%$) \\ \\
   $^{197}$Au
\, \, 77.3 keV &  1.91 ns &
$7/2+$  & $1/2+$ 
& $-42 (29\%)$ & $-0.54 (29\%)$ &
$-2.24 (10\%)$ 
\\ \\
 $^{229}$Th
\, \, 8 eV & $10^3\, $ s &
5/2+ & 3/2+ 
  & $-67\,(13\%)$ &  $-0.82 \, 10^4 (13\%)$  & $-1.19 \, 10^4 (13\%)$ \\ \\
    $^{235}$U
  \, \, 76 eV & 26 m & 
  $7/2-$ & $1/2+$
    & $\sim 100$ &
  $  10^3 $ & $ 10^3$  \\ \\
$^{243}$Am
\, \, 84 keV  & 2.3 ns &
5/2$-$ & 5/2+ 
   & 235 $(25\%)$ & 2.8 $(25\%)$ & 2.6 $(39\%)$  \\ 
\end{tabular}
\end{ruledtabular}
\label{tab:moreelements}
\end{table*}

\appendix

\section{ Nilsson model calculations for 76 eV $^{235}$U transition} \label{Nilsson}

Two low-lying energy transitions are of particular interest due to their high $K$ values and possibility to use high precision atomic spectroscopy methods: the 8 eV $^{229}$Th and the 76 eV $^{235}$U. If these cases are realized with M\"ossbauer spectroscopy, they might bring the method into the UV laser range, in recoil-free resonance. We recently calculated $K_{\alpha}$ for  $^{229}$Th  in \cite{Fadeev2020}. Sensitivity to the quark mass for $^{229}$Th is calculated in the present work using  $K_q=1.45 (K_{\alpha }-1)$. 
Let us now focus on the case of the 76 eV transition in $^{235}$U.

The mean square root charge radius $\langle r^{2} \rangle^{1/2} $ of $^{235}$U is $5.8337(41)$ fm \cite{Angeli2013}. As far as we know, the charge radius for the excited state has not been measured, so we estimate the difference of the charge radii between excited and ground states using the Nilsson deformed oscillator  model \cite{SamThesis}. 
Within the Nilsson model excited and ground states of $^{235}$U differ by the $z$ axis quantum numbers $n_z$ of external neutron: $n_{z,e} = 3$ and $n_{z,g}=4$. For the harmonic oscillator 
\begin{align} \label{rnz}
   \langle z^2 \rangle  = \frac{1}{m_N \omega_z}\left(n_z+\frac{1}{2} \right) \, ,
\end{align}
where $m_N=940$ MeV is the nucleon mass and $\omega_z$ is the harmonic oscillator frequency on the symmetry axis of the deformed nucleus. Substitution of  $n_{z,e} = 3$ and $n_{z,g}=4$ gives 
\begin{align}
   \langle r^{2}_{3} \rangle -  \langle r^{2}_{4} \rangle =  -\frac{1}{m_N \omega_{z}} \,.
\end{align}
The value of $\langle r^{2} \rangle $ is calculated for the total density of all nucleons, therefore, we should divide the result by the number of nucleons $A$. 
\begin{align}
 \Delta  \langle r^{2}\rangle  =  -\frac{1}{A} \frac{(\hbar c)^2}{m_N \omega_{z}} 
 = - \frac{1}{235} \frac{197^2}{940 \times 5}= -0.035 \, \text{fm} \, .
\end{align}
The quadrupole moment of $^{235}$U ground state was measured to be $Q_{lab}= 4.936(6)$ b in the laboratory frame ~\cite{Stone2016}. To calculate the intrinsic quadrupole moment $Q_0$ we use Eq. \eqref{Qlab}
with the nuclear spin $I=7/2$.

As the quadrupole moment of the excited state has not been measured, we estimate the difference of $Q_0$ for the excited and ground states using the Nilsson model.
Quadrupole moment for external neutron is given by 
\begin{align} \label{dq0estimate}
    Q_0 = \frac{1}{A} \left[
    \frac{2}{m_N \omega_z} \left( n_z+\frac{1}{2} \right) -  \frac{2}{m_N \omega_z} \left( N-n_z+1\right)
    \right] \,,
\end{align}
where $N$ is the principle quantum number, which is 6 for the exited state and 7 for the ground state. Equation \eqref{dq0estimate} is a result of using Eq. \eqref{rnz} in the quadrupole moment defined as
\begin{align}
    Q_0 = 2  \langle z^{2}\rangle -  \langle x^{2}+ y^2 \rangle \, .
\end{align}
Inserting numerical values, the change in the intrinsic quadrupole moment of the distribution of $A$ nucleons between the excited and ground states is
\begin{align}
    \Delta Q_0 = \frac{1}{A}  \frac{1}{m_N \omega_z} (6-8) = -0.064 \, \text{fm}^2 \, .
\end{align}
Using these values we obtain $\Delta E_C$=434 keV for the prolate spheroid model and   $\Delta E_C$=191 keV for the constant density model.
These values are few times bigger than $\Delta E_C \sim $ 100 keV in other heavy deformed nuclei. Therefore,  we assume that our Nilsson model calculations overestimate $\Delta E_C$ few times. A conservative estimate is $K_q\approx K_{\alpha} \sim $ 100 keV/76 eV  $ \approx 10^3$.

\section{Inputs for derivation of $K_{\alpha}$} \label{Inputs}
In Table \ref{tab:inputs} we present the inputs we used for each isomeric transition to arrive at the values of $K_{\alpha}$ and $K_q$ in the main text. For $^{57}$Fe, we take the average of three values in \cite{Filatov2009} 
for $d\langle r^2 \rangle$. For $d\langle r^2 \rangle$ of $^{235}$U we take the value derived in the Appendix \ref{Nilsson}. In $^{121}$Sb, $^{127}$I, and $^{129}$I the error on  $d\langle r^2 \rangle$ is unknown.

\begin{table*}[tbp]
\caption{Experimental data used for input. 
The ground state $\langle r^2 \rangle$ values are taken from \cite{Angeli2013}.
$\Delta \langle r^2 \rangle$ referenced individually for each transition, unless taken from M\"ossbauer isomer shifts in \cite{Kalvius74}. Individual values in \cite{Kalvius74}  do not have error bars, only a general ``optimistic"  estimate  of the error $\sim$ 25\% is given. We present \% error in brackets.
$Q_{lab}$ values are from~\cite{Stone2016}. The ratio $Q_{ex}/Q_{gr}$ is taken from the references in \cite{Stone2016}, as many times its uncertainty is smaller than the uncertainty one gets by division of individual values $Q_{ex}$ and $Q_{gr}$. Empty cells of $Q_{ex}/Q_{gr}$ appear when this ratio was obtained by division of individual values $Q_{ex}$ and $Q_{gr}$.
}
\centering
\begin{ruledtabular}
\begin{tabular}{lcccccccccc} 
 &  \multicolumn{2}{c}{$j \pi$} & \multicolumn{2}{c}{} & \multicolumn{3}{c}{$Q_{lab}$} & \\
& ground & excited & $r_{rms}^{gr}$ fm & $d\langle r^2 \rangle 10^{-3}$ fm$^2$ & ground & excited &  $Q_{ex} / Q_{gr}$ \\
\hline \\
$^{57}$Fe
\, \, 14.4 keV & $1/2-$  & $3/2-$  
& {3.7532$\pm$ 0.0017} & $-23.4 \, (4.8 \%)$ \cite{Filatov2009} & 0 & {0.160$\pm 0.008$ } & 
\\ \\ 
  $^{67}$Zn
\, \, 93.3 keV & $5/2-$  & $1/2-$ 
& {3.9530$\pm$ 0.0027}& {18 (22\%)\cite{Aumann1989}}& {0.150$\pm$ 0.015} & 0&
 \\ \\
  $^{83}$Kr
\, \, 9.3 keV & $9/2+$  & $7/2+$ 
  &{4.1871$\pm$ 0.0023} & 6.4 (25\%)& {0.259$\pm$ 0.001}& {0.507$\pm$ 0.003}& {1.958$\pm$ 0.002}
  \\ \\
$^{99}$Ru
\, \, 90 keV & $5/2+$  & $3/2+$  
  & {4.4338$\pm$ 0.0042} & 20 (25\%)& {0.079$\pm$ 0.004}& {0.231$\pm$ 0.013}& {2.93$\pm$ 0.07}
  \\ \\  
    $^{119}$Sn
\, \, 23.9 keV & $1/2+$  & $3/2+$   
& {4.6438$\pm$ 0.0020}& 7.42 (2.3\%)\cite{Filatov2009} & 0 & {$-0.132 \pm$ 0.001} &
 \\ \\
   $^{121}$Sb
\, \, 37.2 keV & 5/2+ & 7/2+ 
& {4.6802$\pm$ 0.0026} & $-0.052$ \cite{Svane2003} & {$-0.543 \pm$ 0.011}& {$-0.727\pm$ 0.016}& {1.34$\pm$ 0.01} 
\\ \\
 $^{125}$Te
\, \, 35.5 keV & 1/2+ & 3/2+  &
 {4.7204$\pm$ 0.0030} &  3.8 (13\%) \cite{Filatov2009} & 0 & {$-0.31\pm$ 0.02}& 
  \\ \\
  $^{127}$I
\, \, 57.6 keV
& $5/2+$  & $7/2+$  &
{4.7500$\pm$ 0.0081} & $-$15.3 \cite{Filatov2009}& {$-0.696\pm$ 0.012}& {$-0.624\pm$ 0.011}& {0.896$\pm$ 0.002} \\ \\
  $^{129}$I
\, \, 27.8 keV & 7/2+ & 5/2+ 
& {4.74$\pm$ 0.01} & 19.3 \cite{Filatov2009} & {$-$0.488$\pm$ 0.008}& {$-0.604\pm$ 0.010}& {1.2385$\pm$ \
0.0011} \\ \\
  $^{149}$Sm
\, \, 22.5 keV & $7/2-$  & $5/2-$ 
  & {5.0134$\pm$ 0.0035}& 1.3 (25\%)& {0.078$\pm$ 0.008}& {1.01$\pm$ 0.09} &
   \\ \\
  $^{151}$Eu \, 22 keV & 5/2+ &7/2+ 
& {5.0522 $\pm$ 0.0046} & {24.8 (29\%)\cite{Tanaka1984Eu}}& {0.903 $\pm$ 0.010}& {1.28 $\pm$ 0.02} &
\\ \\
 $^{153}$Eu
\, \, 83.4 keV & $5/2+$  & $7/2+$  
  &{5.1115$\pm$ 0.0062}& $-$2.7 (25\%)& {2.41$\pm$ 0.02}& {0.44$\pm$ 0.02}& 
  \\ \\
$^{153}$Eu 103 keV & 5/2+ & 3/2+
& {5.1115 $\pm$ 0.0062} & $-85$(15\%) \cite{Hess1969}& {2.41 $\pm$ 0.02}& {1.253$\pm$ 0.012}&
{0.52$\pm$ 0.003} \\ \\
   $^{155}$Gd
  \, 86.5 keV & $3/2-$ & $5/2+$ 
  & {5.1319$\pm$ 0.0041} & $-$5.6 (25\%)& {1.27$\pm$ 0.03}& {0.110$\pm$ 0.008}& {0.087$\pm$ 0.006}
  \\ \\
$^{155}$Gd 105 keV & $3/2-$ & 3/2+ 
& {5.1319$\pm$ 0.0041} & $-7.7 \left(25\%\right)$ & {1.27$\pm$ 0.03}&  {1.27$\pm$ 0.05} & {1$\pm$ 0.03} \\ \\
$^{157}$Gd
\, 64 keV & 3/2$-$ & 5/2+ 
&{5.1449 $\pm$ 0.0042} & 14.3 $\left(25\% \right)$ & {1.35 $\pm$ 0.03} & {2.43 $\pm$ 0.07}& {1.80 $\pm$ 0.03} \\ \\
    $^{158}$Ho
\, \, 67.2 keV &  $5+$  & $2- $ 
  & {5.1571$\pm$ 0.0316}& & {4.2$\pm$ 0.4}& {1.66$\pm$ 0.17}&
  \\ \\
     $^{160}$Ho
\, \, 60 keV & $5+$  & $2- $
  & {5.1662$\pm$ 0.0315}& & {4.0$\pm$ 0.2}& {1.83$\pm$ 0.17}&
  \\ \\
$^{161}$Dy
\, \, 25.7 keV & 5/2+ & $5/2-$ 
  & {5.1962$\pm$ 0.0459}& 7.4 (25\%)& {2.51$\pm$ 0.02}& {2.51$\pm$ 0.02}&
   \\ \\  
 $^{161}$Dy
\, \, 43.8 keV & 5/2+ & 7/2+ 
  & {5.1962$\pm$ 0.0459} & $-$1.6 (25\%)& {2.51$\pm$ 0.02}& {0.53$\pm$ 0.13}& {0.21 $\pm$ 0.05} \\ \\
$^{161}$Dy
\, 75 keV  & 5/2+ & 3/2$-$ 
 & {5.1962$\pm$ 0.0459} & $7.9 (25\%)$ & {2.51$\pm$ 0.02} & {1.45$\pm$ 0.06} & {0.577$\pm$ 0.025} \\ \\
$^{165}$Ho
\, \, 95 keV & $7/2-$  & $9/2- $ 
  & {5.2022$\pm$ 0.0312}& & {3.58$\pm$ 0.02}& {3.52$\pm$ 0.04}& {0.983$\pm$ 0.008}   \\ \\
  $^{179}$Hf
  \, \, 123 keV & $9/2+$ & $11/2+ $
  & {5.3408$\pm$ 0.0031}& & {3.79$\pm$ 0.03}& {1.88$\pm$ 0.03}& 
  \\ \\
  $^{181}$Ta
\, \,  6 keV & 7/2+ & 9/2$-$ 
 & {5.3507$\pm$ 0.0034} & $-43 (25\%)$ & {3.17$\pm$ 0.02} & {3.59$\pm$ 0.02} & {1.1315$\pm$ 0.0002} \\ \\
      $^{197}$Au
\, \, 77.3 keV &  $3/2+$  & $1/2+$ 
& {5.4371$\pm$ 0.0038} & {8.28}(29\%)  \cite{Filatov2009} &  {0.547$\pm$ 0.016} & 0 &
\\ \\
$^{229}$Th
\, \, 8 eV & 5/2+ & 3/2+ 
  & 5.7557 $\pm$ 0.0143 & {10.5 (12\%)\cite{Safronova2018}}& 3.13 $\pm$ 0.03& {1.74$\pm$ 0.06} & 0.555$\pm$ 0.019 \cite{Thielking2018}
  \\ \\ 
     $^{233}$U
  \, \, 40 keV & $5/2+$ & $7/2+$ 
  & {5.8203$\pm$ 0.0049} & & {3.663$\pm$ 0.008}& {0.64$\pm$ 0.03} &
  \\ \\ 
  $^{235}$U
\, \, 46 keV & $7/2-$ & $9/2-$ 
  & {5.8337$\pm$ 0.0041} &  & {4.936$\pm$ 0.006}& {1.87$\pm$ 0.03} &
  \\ \\
    $^{235}$U
  \, \, 76 eV & $7/2-$ & $1/2+$
    & {5.8337 $\pm$ 0.0041} & $-$35 \cite{appA} & {4.936$\pm$ 0.006} & 0 & &  \\ \\
  $^{243}$Am
\, 84 keV  & 5/2$-$ & 5/2+ 
   & {5.9048, 0.0035} & $-$42 (25\%) & {4.32$\pm$ 0.06} & {4.2$\pm$ 0.2} & {0.97$\pm$ 0.05} \\
\end{tabular}
\end{ruledtabular}
\label{tab:inputs}
\end{table*}

\FloatBarrier
\bibliographystyle{prsty}

\end{document}